\documentclass[sigconf]{acmart}

\usepackage{makecell}
\usepackage{multirow}
\usepackage{longtable}
\usepackage{graphicx}
\usepackage{caption}
\usepackage{subcaption}
\usepackage{tabularx}
\usepackage{soul}
\usepackage{float}
\usepackage{url}
\usepackage{booktabs}
\usepackage{colortbl}

\usepackage{enumitem}

\usepackage{tcolorbox}
\tcbuselibrary{skins,breakable}

\newif\ifredact
\redactfalse  

\newcommand{\pci}{parent--child interaction}

\newif\ifcomment
\commenttrue   
\ifcomment
  \newcommand{\missing}[1]{\textcolor{red}{~#1}}
  \newcommand{\wei}[1]{~\sethlcolor{cyan!40}\hl{[Weiyan: #1]}}
  \newcommand{\ken}[1]{~\sethlcolor{yellow!40}\hl{[Kenny: #1]}}
  \newcommand{\rrev}[3]{\textcolor{blue}{[RevID #1] #2: #3}} 
\else
  \newcommand{\missing}[1]{}
  \newcommand{\wei}[1]{}
  \newcommand{\ken}[1]{}
  \newcommand{\rrev}[3]{}
\fi

\AtBeginDocument{%
  }

\copyrightyear{2026}
\acmYear{2026}
\setcopyright{cc}
\setcctype{by}
\acmConference[CHI 2026 BiAlign Workshop]{CHI 2026 BiAlign Workshop}{April 13--17, 2026}{Barcelona, Spain}

\begin{document}

\title[Human-MLLM Alignment in Early Developmental Communities]{More Than 1v1: Human-AI Alignment in Early Developmental Communities with Multimodal LLMs}

\author{Weiyan Shi}
\email{weiyanshi6@gmail.com}
\orcid{0009-0001-6035-9678}
\affiliation{
  \institution{Singapore University of Technology and Design}
  \city{Singapore}
  \country{Singapore}
}

\author{Kenny Tsu Wei Choo}
\email{kennytwchoo@gmail.com}
\orcid{0000-0003-3845-9143}
\affiliation{
  \institution{Singapore University of Technology and Design}
  \city{Singapore}
  \country{Singapore}
}




\begin{abstract}
In early developmental contexts, particularly in parent–child interaction analysis, alignment involves families and professionals—such as speech-language pathologists (SLPs)—who interpret children’s everyday interactions from different roles. When multimodal large language models (MLLMs) are introduced to support this process, alignment becomes a question of how authority, responsibility, and emotional risk are distributed across stakeholders.
Through a three-part study with five families and three SLPs, we trace how MLLM-generated outputs move from expert-facing analysis to parent-facing feedback. We propose layered community alignment: grounding representations in expert-aligned structures, mediating translation through professional guardrails, and enabling family-level adaptation within those boundaries. We argue that alignment in developmental settings should be treated as a community-governed process rather than an individual optimisation problem.
\end{abstract}

\maketitle

\section{Introduction}

Human–AI alignment~\cite{shen2024valuecompass} is often framed as a problem of matching model outputs to the preferences, values, or opinions of individual users~\cite{shankar2024validates} or specific stakeholder groups~\cite{shi2025aligning}. While such approaches have advanced techniques for persona modelling, value representation, and demographic mirroring, they largely assume that alignment is a model-to-user optimisation problem: the system should produce outputs that best correspond to the expectations of a target audience.

However, in many real-world domains, alignment is not negotiated between a model and a single user. It unfolds within communities where multiple stakeholders interpret the same underlying material under asymmetric roles and responsibilities. In early developmental contexts, for example, \pci{} videos recorded at home may be reviewed by both families and speech-language pathologists (SLPs)~\cite{snodgrass2017telepractice}. These actors share a developmental goal, yet operate under distinct epistemic norms, emotional sensitivities, and professional accountabilities. 

When multimodal large language models (MLLMs) are introduced to analyse such shared data, a fundamental tension emerges. Should the system optimise for clinical accuracy, emotional safety, parental empowerment, or professional authority? Directly aligning model outputs to parents may risk bypassing expert judgement; aligning solely to expert language may render outputs psychologically inappropriate for families. In such settings, alignment is not merely about ``sounding right'' to a demographic group—it becomes a safety and governance challenge concerning who mediates interpretation, how responsibility is distributed, and how authority is preserved.
Existing work on alignment as opinion mirroring~\cite{do2025aligning, liu2025alignment} provides useful techniques for modelling population-level perspectives, but offers limited guidance for scenarios where stakeholders are structurally asymmetric and co-interpret the same artefact. From an HCI perspective, this raises a broader concern: alignment in community settings cannot be reduced to preference matching. It must instead account for layered interpretation, mediated translation, and context-sensitive adaptation.

In this paper, we argue that alignment in early developmental domains should be organised as a layered community process rather than treated as a single optimisation objective. Specifically, we propose that responsible deployment of MLLMs in \pci{} contexts requires three structurally distinct layers: (1) expert-aligned behavioural representation, (2) expert-mediated reframing for caregiver use, and (3) family-level contextual personalisation within professional guardrails. We refer to this as \emph{layered community alignment}.

To ground this position, we draw on a three-part study involving families (N = 5) and SLPs (N = 3) using naturalistic home-recorded \pci{} videos. We use the study as an empirical lens to surface tensions in how expert-facing analyses are translated, constrained, and adapted for family use. Through this lens, we examine the following question:

\textbf{RQ: How should MLLMs be aligned in community settings where multiple stakeholders with asymmetric roles interpret the same child’s everyday interactions?}

In this workshop paper, we advance a reframing of human–AI alignment in early developmental communities:

First, we propose \textbf{layered community alignment}, showing that human-AI alignment in early developmental contexts unfolds across three layers: expert-facing representation, expert-mediated translation, and family-level adaptation.

Second, we identify three recurring \textbf{alignment tensions} across these layers—between representational alignment and professional authority, clinical precision and emotional safeguarding, and standardisation and contextual adaptation—framing alignment as a negotiated, multi-stakeholder process.

\section{Empirical Lens: Probing Alignment Tensions}

To ground our position in real-world practice, we conducted a three-part exploratory study examining how MLLM-generated analyses of \pci{} are interpreted, mediated, and adapted across stakeholder roles. 

Rather than serving as a formal validation of system performance, this study functioned as an empirical probe to surface structural tensions in community-based alignment.

\subsection{Design Overview}

\begin{figure}[htbp]
    \centering
    \includegraphics[width=\linewidth]{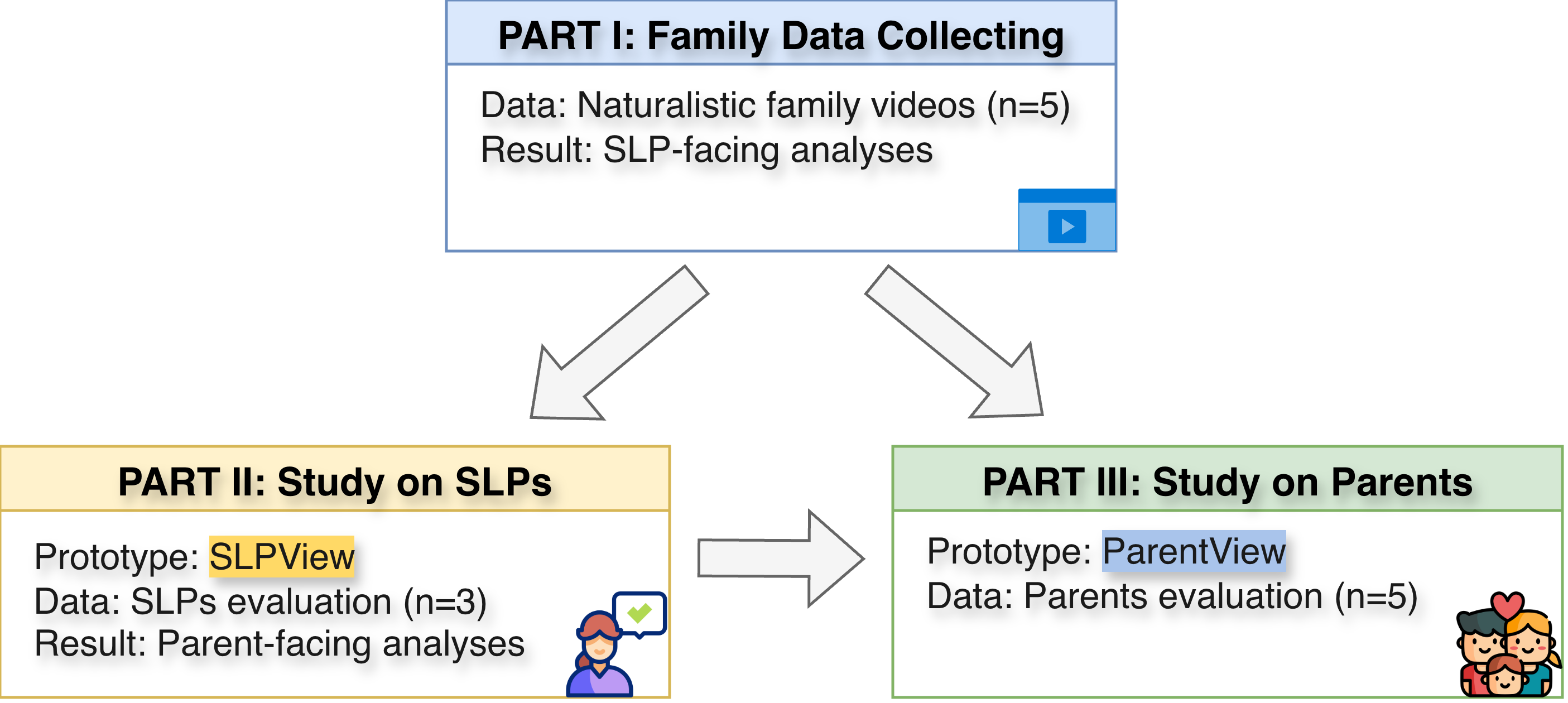}
    \caption{Overview of the three-part study. 
    \textbf{Part I} collected naturalistic \pci{} videos and generated expert-aligned SLP-facing analyses. 
    \textbf{Part II} involved SLP evaluation and steering of expert outputs. 
    \textbf{Part III} evaluated a parent-facing prototype grounded in practitioner-informed guidance.}
    \label{fig:overview}
\end{figure}

\begin{figure*}[htbp]
    \centering
    \includegraphics[width=0.9\linewidth]{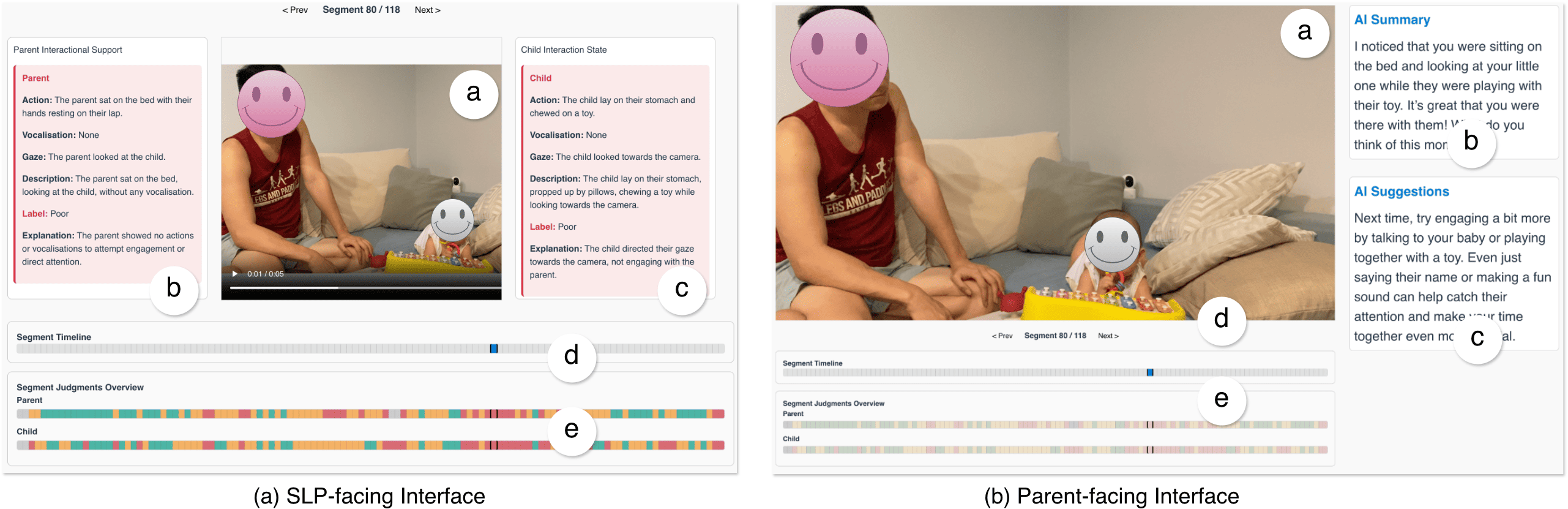}
    \caption{
    \textbf{System Overview.}
    Both interfaces are grounded in the same segment-level MLLM analysis of parent–child interaction (\pci{}) videos.
    \textbf{SLPView (left)} provides structured expert-oriented behavioural assessment with quality labels and aligned timelines.
    \textbf{ParentView (right)} reframes this analysis into descriptive summaries and supportive suggestions, removing evaluative framing while preserving segment-level continuity.
    }
    \label{fig:system-overview}
\end{figure*}

The study unfolded in three parts (Figure~\ref{fig:overview}):

\begin{enumerate}
    \item \textbf{Part I: Family Data Grounding (N = 5 Families).} We recruited parents with children under three years old and selected five families (F1--F5) to contribute self-recorded, naturalistic \pci{} videos captured in everyday home environments. These videos provided real-world interaction material for subsequent expert review and alignment analysis.
    
    \item \textbf{Part II: SLP Evaluation and Steering Design (N = 3 SLPs).} We recruited three certified SLPs (7--9 years of paediatric experience) to evaluate MLLM-generated SLP-facing analyses of the family videos. Using an SLP-facing prototype (Figure~\ref{fig:system-overview}a), SLPs reviewed structured segment-level outputs produced by an existing expert-aligned two-stage prompting pipeline~\cite{shi2025aligning,shi2025towards}. Stage 1 generated behavioural observations (action, vocalisation, gaze), and Stage 2 produced role-specific interaction judgements with brief explanations. Data collection included a pre-interview on practice related to \pci{}, think-aloud prototype use, a post-task questionnaire assessing human-centred AI dimensions (alignment, accuracy, relevance, safety, transparency, collaboration)~\cite{shneiderman2020bridging}, and a post-interview focusing on steering guidelines for translating expert-facing analyses into parent-facing outputs and on collaboration potential with families.
    
    \item \textbf{Part III: Parent-Facing Evaluation (N = 5 Families).} Based on SLP steering guidance, we developed a parent-facing prototype (Figure~\ref{fig:system-overview}b) presenting supportive summaries and suggestions derived from the SLP-facing outputs. Parent-facing analyses were generated using a retrieval-augmented generation pipeline grounded in practitioner-authored parent-coaching texts (\textit{It Takes Two to Talk}~\cite{pepper2004talk}, \textit{More Than Words}~\cite{sussman1999words}, \textit{The Floortime Manual}~\cite{dir_floortime}, and \textit{Strategies for Building Language and Communication}~\cite{CPAS2020strategies}), guided by SLP-informed steering prompts.
\end{enumerate}

\subsection{Analytic Approach}

All sessions were audio-recorded and transcribed. Our analysis focused on identifying recurring points of friction, negotiation, and reinterpretation across stakeholder roles. 

For SLP data, we examined how experts interpreted the strengths and limits of MLLM-generated analyses, how they articulated boundaries for translating clinical representations into parent-facing outputs, and how they positioned the system within existing professional–family collaboration workflows.

For parent data, we examined how caregivers interpreted the adequacy and relevance of MLLM-generated feedback in relation to lived interaction experiences, and how they envisioned the system’s role in ongoing collaboration with professionals.

\section{Tracing Alignment Tensions Across Layers}

Our analysis reveals how alignment shifts across three structurally distinct layers: expert-facing representation, expert-mediated translation, and family-level adaptation. Each layer exposes a different boundary in how MLLM outputs travel across stakeholder roles in \pci{} contexts.

\subsection{Layer 1: Expert-Readable Representation — Alignment as Structured Visibility}

At the foundational layer, MLLM outputs were structured to mirror expert observational logic. The system separated behavioural evidence (action, vocalisation, gaze) from interaction-level interpretation, reflecting professional reasoning patterns in speech-language pathology~\cite{podder2023soap}.
SLPs recognised this structure as legible and familiar. The analytic categories aligned with how they document joint attention and parent support in practice. However, experts consistently distinguished between representational alignment and clinical judgement. While the model could externalise observable cues in expert-readable form, it could not situate these cues within longitudinal developmental trajectories, case history, or therapeutic reasoning.

This layer surfaces a first tension: alignment at the level of structure does not confer professional authority. The model can reproduce the language of expertise, yet responsibility and interpretive accountability remain human.

\subsection{Layer 2: Expert-Mediated Translation — Alignment as Emotional Safeguarding}

When outputs moved from expert-facing analysis to parent-facing reflection, a different boundary became visible. SLPs expressed concern that directly exposing diagnostic-style labels or categorical judgements (e.g., “Poor”) could be psychologically harmful or misinterpreted without clinical framing.
Three SLPs articulated constraints for reframing outputs: softening evaluative tone, filtering categorical labels, and translating therapeutic goals into everyday interaction suggestions.

This layer exposes a second tension: clinical precision and emotional safety are distinct alignment objectives. What is informative and appropriate for professionals may not be suitable for caregivers. Alignment therefore requires mediation that redistributes interpretive authority rather than simply rephrasing model outputs.

Here, expert steering operates as ethical infrastructure, defining boundaries for what should be surfaced, softened, or withheld.

\subsection{Layer 3: Family-Level Adaptation — Alignment as Contextual Negotiation}

Even after expert-mediated reframing addressed safety concerns, parents identified another boundary: contextual fit. Caregivers described some outputs as accurate yet “textbook-like,” requesting adaptation to temperament, developmental stage, mood, and daily routines.
This demand did not reject expert-defined guardrails. Instead, it highlighted that alignment does not conclude at professional mediation. Families sought flexibility within boundaries, not outside them.

This layer surfaces a third tension: standardisation versus contextualisation. While expert guardrails ensure safety and coherence, rigid framing risks overlooking lived variability. Alignment in community settings therefore extends beyond representation and translation—it involves negotiated adaptation within professional constraints.
\section{Conclusion and Discussion}

This work positions alignment in early developmental contexts not as a single modelling objective, but as a community-governed process that unfolds across representation, mediation, and contextual negotiation. By tracing how MLLM outputs move between SLPs and families, we surface structural tensions that extend beyond \pci{} and raise broader questions about how multimodal AI systems should participate in professional–lay ecosystems.
In sensitive developmental domains, AI systems do not merely generate text; they intervene in interpretive relationships between experts and caregivers. Alignment therefore becomes not only a technical problem, but a governance question concerning authority, responsibility, and emotional consequence.

\subsection{Alignment and Professional Authority}

At the expert-facing layer, structured behavioural representation enhances analytic visibility. By separating gaze, vocalisation, and action into segment-level breakdowns, the system mirrors professional documentation practices. This can increase efficiency and make implicit observation more explicit.

However, structured visibility also introduces a subtle risk: the appearance of expertise without accountability. When a model adopts the structure and vocabulary of clinical reasoning, it may be perceived as possessing interpretive authority. Yet it lacks access to longitudinal history, therapeutic rapport, and developmental nuance.

This raises a foundational governance question:
\emph{When AI systems reproduce expert reasoning structures, how do we prevent them from being perceived as clinical decision-makers?}

If structured outputs are treated as quasi-diagnostic artefacts, the boundary between support tool and authoritative evaluator may blur. In developmental contexts, where labels carry long-term psychological implications, this distinction is critical. Alignment at the representational level must therefore be explicitly framed as assistance, not judgement.

\subsection{Alignment and Emotional Risk}

The transition from expert-facing outputs to parent-facing feedback revealed a second tension: the divergence between clinical precision and emotional safety.

Diagnostic-style categories, colour-coded signals, or evaluative labels may be clinically meaningful but emotionally destabilising. Parents may internalise such signals as judgements of their caregiving competence. Conversely, excessive softening may obscure important developmental concerns, delaying intervention.

This tension is not merely linguistic; it concerns the ethical distribution of interpretive power. 

\emph{Who decides what level of transparency is appropriate?}
\emph{How do we balance the caregiver’s right to know with the responsibility to prevent harm?}

Expert-mediated steering in our study functioned as an ethical filter. However, this introduces another risk: paternalism. If experts systematically mediate or withhold interpretive signals, families may be shielded from information they deem relevant. 

Designing alignment in such contexts requires making these filtering logics visible and contestable rather than invisible defaults embedded in prompts.

\subsection{Alignment, Personalisation, and Epistemic Drift}

At the family level, contextual adaptation becomes central. Parents sought outputs that reflected temperament, mood, fatigue, and daily routines. This highlights the importance of situational grounding in real-world caregiving.

Yet personalisation also introduces epistemic drift. If outputs become overly tailored to family preferences, they may gradually deviate from professional standards. 

This raises a structural question:
\emph{How can systems support contextual flexibility without eroding clinical coherence?}

Layered alignment suggests that personalisation should operate within expert-defined guardrails. However, scaling such guardrails across diverse families, cultures, and developmental conditions remains an open challenge. 

Automation bias further complicates this dynamic. Parents may over-trust structured feedback, particularly when presented with timelines, scores, or systematic breakdowns. Even when outputs are framed as supportive suggestions, their structured appearance may implicitly signal authority.

\subsection{Alignment as Distributed Responsibility}

Across these layers, alignment emerges as a redistribution of responsibility rather than a refinement of prompts. The model structures visibility. Experts define mediation boundaries. Families negotiate contextual adaptation.

This distribution prompts further workshop questions:

\begin{itemize}
    \item Should MLLMs in developmental domains ever operate without professional mediation?
    \item How should liability be handled if AI-generated interpretations influence caregiver decisions?
    \item What mechanisms can make expert guardrails transparent rather than opaque?
    \item How can layered alignment scale without concentrating interpretive power in either models or experts?
\end{itemize}

Although our empirical lens focuses on \pci{}, similar asymmetries exist in mental health support, special education, rehabilitation, and elder care. In these domains, AI systems interact with vulnerable populations and established professional hierarchies. Alignment must therefore account not only for output quality, but for how authority and accountability are structured.

We propose layered community alignment not as a completed framework, but as a conceptual scaffold for rethinking how multimodal AI systems participate in expert–family ecosystems. Future research must move beyond optimising model responses and instead design infrastructures that clarify interpretive roles, safeguard emotional well-being, and maintain professional coherence while allowing contextual flexibility.

\subsection{From Community Alignment to Societal Infrastructure}

While our analysis centres on SLP–family ecosystems, layered alignment also has implications beyond immediate community boundaries. In practice, MLLM-based systems for developmental analysis are unlikely to remain confined to small clinical settings. They may be integrated into telehealth platforms, parenting applications, insurance-supported interventions, or large-scale educational infrastructures.

At this broader scale, alignment becomes not only a matter of expert mediation, but of institutional design.

Several societal-level questions emerge:

\begin{itemize}
    \item Who is authorised to deploy AI-generated developmental analyses at scale?
    \item Should expert mediation be mandatory, optional, or context-dependent?
    \item How should transparency be structured so that families understand when outputs are AI-generated, expert-filtered, or directly model-produced?
    \item What regulatory safeguards are necessary to prevent misinterpretation, over-reliance, or inappropriate substitution of professional care?
\end{itemize}

If layered alignment is treated merely as a design choice, it may be bypassed in commercial deployments prioritising accessibility or scalability. However, in sensitive developmental domains, bypassing expert mediation could amplify emotional harm, misinformation, or inequitable access to care.

This suggests that layered alignment may need to be formalised not only as an interaction pattern, but as infrastructural policy. Mechanisms such as professional certification layers, audit trails of AI-generated interpretations, and explicit disclosure of mediation processes could help maintain accountability at scale.

In this sense, community alignment scales into societal governance. Designing multimodal AI for early development requires attention not only to how outputs are framed, but to how interpretive authority is distributed across platforms, institutions, and regulatory systems.

\bibliographystyle{ACM-Reference-Format}
\bibliography{main}

\appendix
\end{document}